\documentclass[conference]{IEEEtran}
\IEEEoverridecommandlockouts
\usepackage{cite}
\usepackage{url}
\usepackage{xcolor}
\usepackage{comment}
\usepackage{array}
\usepackage{amsmath,amssymb,amsfonts}
\usepackage{graphicx}
\usepackage{textcomp}
\usepackage{xcolor}
\usepackage{algorithm}
\usepackage{algpseudocode}
\usepackage{mathptmx}
\usepackage{color,soul}
\usepackage{multirow}
\usepackage{booktabs}
\usepackage{amsfonts}
\def\BibTeX{{\rm B\kern-.05em{\sc i\kern-.025em b}\kern-.08em
    T\kern-.1667em\lower.7ex\hbox{E}\kern-.125emX}}
\begin{document}
\title{High Impedance Fault Detection Through Quasi-Static State Estimation: A Parameter Error Modeling Approach\\

}
\author{
\IEEEauthorblockN{Austin Cooper}
\IEEEauthorblockA{\textit{Dept. of ECE} \\
\textit{University of Florida}\\
Gainesville, FL, USA.\\austin.cooper@ufl.edu}
\and

\IEEEauthorblockN{Arturo Bretas}
\IEEEauthorblockA{\textit{Distributed Systems Group}\\
\textit{Pacific Northwest National Laboratory}\\
Richland, WA, USA.\\arturo.bretas@pnnl.gov}
\and

\IEEEauthorblockN{Sean Meyn}
\IEEEauthorblockA{\textit{Dept. of ECE} \\
\textit{University of Florida}\\
Gainesville, FL, USA.\\meyn@ece.ufl.edu}
\and

\IEEEauthorblockN{Newton G. Bretas}
\IEEEauthorblockA{\textit{Dept. of ECE} \\
\textit{University of São Paulo}\\
São Carlos, SP, BRA.\\ngbretas@sc.usp.br}
}

\maketitle

\begin{abstract}
 This paper presents a model for detecting high-impedance faults (HIFs) using parameter error modeling and a two-step per-phase weighted least squares state estimation (SE) process. The proposed scheme leverages the use of phasor measurement units and synthetic measurements to identify per-phase power flow and injection measurements which indicate a parameter error through $\chi^2$ Hypothesis Testing applied to the composed measurement error (CME). Although current and voltage waveforms are commonly analyzed for high-impedance fault detection, wide-area power flow and injection measurements, which are already inherent to the SE process, also show promise for real-world high-impedance fault detection applications. The error distributions after detection share the measurement function error spread observed in proven parameter error diagnostics and can be applied to HIF identification. Further, this error spread related to the HIF will be clearly discerned from measurement error. Case studies are performed on the 33-Bus Distribution System in Simulink.
\end{abstract}

\begin{IEEEkeywords}
High impedance fault, Fault detection techniques, State Estimation, Parameter error detection, Wide Area Measurements
\end{IEEEkeywords}

\section{Introduction}\label{sec:intro}
As the complexity of power distribution systems (PDS) and distributed energy resource (DER) penetration increases, so too does the need for improving diagnostics and reliability. One such problem in this area is high-impedance faults (HIFs). Despite their ubiquity in PDS, HIFs persist without generalized solution for detection due to many characteristics of the HIF itself, including nonlinearity of voltage and current measurements at the monitored line, electrical arc transient spikes, and the inverse relationship between impedance and measurable fault location current \cite{Incipent}.

Various strategies have been developed to detect and identify HIFs \cite{Review}. A popular domain is signal processing, in which methods based on wavelet transform (WT) \cite{wavelet} are dominant. These techniques use time and frequency information for various power system disturbances, including HIFs. Artificial neural networks (ANNs) have been used, alone and in tandem with WT \cite{ANN}, to decompose and extract features of current and/or voltage waveforms of a monitored line under HIF. Essentially, these techniques identify special waveform features indicative of HIFs and aim to discern them from other power system disturbances, such as switching events. An estimation approach using wide-area measurements presented in \cite{Resapproach} uses normalized residuals, considering only the detectable error component for general fault detection. A PDS model-based approach in \cite{PDS} identifies HIFs based on impulse signal injections and line impedance analysis. While promising, this method is limited to medium voltage system transformer configurations.

The proposed model is based on power system state estimation (SE), which has become consolidated as a necessary procedure for real-time power energy system management \cite{bretas2021cyber}. 
Pertinent to this work are SE gross error (GE) analytics, including the detection and identification of errors in measurement data—either nefarious or equipment-based. The proposed HIF detection model in this work is based on the GE analytics of parameter errors \cite{ParameterProof}, during which network data such as line impedance values do not match those of the PDS due to physical disturbance, system error, or cyber-attack. The proposed work requires minimum modifications to existing SE algorithms. The model detects HIFs regardless of changes in power flow direction due to DERs, as this information is included in the measurement vector.  Further, this work makes no assumptions about system topology or configuration, which are treated as logical inputs to the SE, making it practical for generalized implementation and use in
industry.

The idea for this work is to combine established parameter error detection methods, per-phase weighted least squares SE, and a two-step SE approach to detect HIFs from a wide-area system perspective and, after uncovering information about the error distributions and eliminating the possibility of measurement error, identify the lateral line on which the HIF occurred based on the measurement set fed to the SE process. No training or pattern recognition is required.

The three specific contributions of this paper towards the state-of-the-art are the following:
\begin{itemize}
    \item A formal model for HIF detection based on SE.
    \item HIF detection that requires no assumptions of a fault or system disturbance having already occurred.
    \item A method that is independent of knowledge of the HIF condition with respect to time (e.g., before, during, or after the build-up and shoulder characteristic).
\end{itemize}

\section{Formal Model}\label{sec:formalmodel}
\subsection{HIF Modeling}\label{sec:hifmodel}

HIFs exhibit special characteristics which differentiate them from typical short circuit faults \cite{Review}. These include:
\begin{enumerate}
    \item \textbf{Low-magnitude fault currents:}  HIFs are difficult to discern from normal load changing conditions.
    \item \textbf{Intermittent arcing:} the current waveform does not exhibit a steady state waveform, but rather alternates between conduction and non-conduction cycles.
    \item \textbf{Asymmetry and randomness:} the difference between breakdown voltages causes differences in shape and magnitude between the negative and positive half-cycles of the fault current. 
    \item \textbf{Nonlinearity:}  arcing causes non-linearity between the voltage and current waveforms at the HIF.
    \item \textbf{Build-up and Shoulder:} an overall gradual increase in fault current until reaching a steady state condition for several cycles.
\end{enumerate}

\begin{figure}
\includegraphics[width=6.5072cm,height = 3cm,scale = 1]{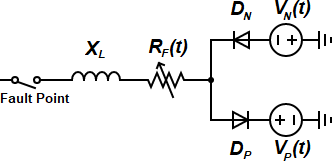}
\centering
\caption{HIF Model \cite{RFT}}
\label{HIF Model}
\end{figure}

 To capture each fault current characteristic, a HIF model was developed, illustrated in Fig. \ref{HIF Model}. To simulate the build-up and shoulder characteristic, a time-varying resistance $R_f(t)$ was implemented using polynomial regression approximation \cite{RFT}. To test the detection capabilities of the proposed work, the variation range of the $R_f(t)$ waveforms was restricted such that the fault current remains as low as detectably possible with respect to full load current of the faulted phase.

Two antiparallel diodes $D_p$ and $D_n$ and two nonequal voltage sources $V_p(t)$ and $V_n(t)$ serve to simulate nonlinearity and the HIF arcing characteristic \cite{MM}. Intermittency was simulated by selecting a portion of the $R_f(t)$ waveform and inserting a high impedance impulse such that the evolving HIF current waveform is temporarily extinguished before resuming.

\subsection{Innovation-Based State Estimation}\label{sec:IISE}
Classical quasi-static state estimation implements a Weighted Least Squares (WLS) method \cite{bretas2021cyber}. A power system with $n$ buses and $d$ measurements is modeled as a set of non-linear algebraic equations in the measurement model:
\begin{equation}
    \mathbf{z} = h(\mathbf{x}) + \mathbf{e}
\label{eq:SE}\end{equation}
where $\mathbf{z}\in\mathbb{R}^{1 \times d}$ is the measurement vector, $\mathbf{x}\in\mathbb{R}^{1 \times N}$ is the vector of state variables, $h:\mathbb{R}^{1 \times N}\rightarrow\mathbb{R}^{1 \times d}$ is a continuous nonlinear differentiable function, and $\mathbf{e}\in\mathbb{R}^{1 \times d}$ is the measurement error vector.  The measurement error, $e_i$ is assumed to have a Gaussian probability distribution with zero mean and standard deviation $\sigma_i$.  $d$ is the number of measurements and $N=2n-1$ is the number of unknown state variables, namely the complex bus voltages.

In the classical WLS method, the optimal state vector estimate in \eqref{eq:SE} is found by minimizing the weighted norm of the residual, represented by the cost function $J(\mathbf{x})$:
\begin{equation} 
\label{eq:JSE}
J(\mathbf{x})=\Vert \mathbf{z}-h(\mathbf{x})\Vert _{\mathbf{R}^{-1}}^{2}=[\mathbf{z}-h(\mathbf{x})]^{T}\mathbf{R}^{-1}[\mathbf{z}-h(\mathbf{x})] 
\end{equation}
where $\mathbf{R}$ is the covariance matrix of the measurements. For this work, a two-step approach will be adopted for the GE analytic process \cite{twostep}. In the first step, all measurements are weighted equally proportional to the measurement magnitude, after which GE analytics are performed. After GE processing, the SE is repeated—this time with meter precision values per state-of-the-art methodologies, which consider a specific standard deviation value for each measurement type \cite{monticelli}.
The solution of \eqref{eq:JSE} is obtained through the iterative Newton-Raphson method. The linearized form of \eqref{eq:SE} becomes:

\begin{equation} \Delta \mathbf{z}=\mathbf{H}\Delta \mathbf{x}+\mathbf{e} \label{SElin} \end{equation}
where $\mathbf{H}=\frac{\partial h}{\partial \mathbf{x}}$ is the Jacobian matrix of $h$ at the current state estimate $\mathbf{x}^*$, $\Delta \mathbf{z}=\mathbf{z}-h(\mathbf{x}^*)=\mathbf{z}-\mathbf{z}^*$ is the measurement vector correction and $\Delta \mathbf{x}=\mathbf{x}-\mathbf{x}^*$ is the state vector correction.  The WLS solution can be understood geometrically \cite{geom} as the projection of $\Delta \mathbf{z}$ onto the Jacobian space $\mathfrak{R}$($H$) by a linear projection matrix $\mathbf{K}$, i.e. $\Delta\hat{\mathbf{z}}=\mathbf{K}\Delta\mathbf{z}$.  


\begin{equation}\label{eq:P}
\mathbf{K} = \mathbf{H}(\mathbf{H}^{T}\mathbf{R}^{-1}\mathbf{H})^{-1}\mathbf{H}^{T}\mathbf{R}^{-1}.
\end{equation}

By decomposing the measurement vector space into a direct sum of $\mathfrak{R}$($H$) and $\mathfrak{R}($H$)^{\perp}$, it is then possible to decompose the measurement error vector $\mathbf{e}$ into two components:
\begin{equation} \label{eq:e}
\mathbf{e} = \underbrace{\mathbf{K}\mathbf{e}}_{\mathbf{e_U}} + \underbrace{(I-\mathbf{K})\mathbf{e}}_{\mathbf{e_D}}.
\end{equation}

The component $\mathbf{e_D}$ is the detectable error, which is equivalent to the residual in the classical WLS model. The component $\mathbf{e_U}$ is the undetectable or masked component of the error. 
To quantify the impact of the undetectable error, the measurement Innovation Index ($II$) is introduced \cite{II}:
\begin{equation}\label{eq:10}
{II}_{i} = \frac{\Vert{e^i_D}\Vert_{\mathbf{R}^{-1}}}{\Vert{e^i_U}\Vert_{\mathbf{R}^{-1}}} = \frac{\sqrt{1-K_{ii}}}{\sqrt{K_{ii}}}.
\end{equation}

Low Innovation Index means that a large component of the error is not reflected in the residual alone. The residual will therefore be very small even if there is a GE. From (\ref{eq:e})  and (\ref{eq:10}), the composed measurement error ($CME$) can be expressed in terms of the residual and the innovation index, after which the normalized $CME^N$ is obtained:
\begin{equation}\label{eq:11}
CME_i = r_i\left(\sqrt{1+\frac{1}{{II_i}^2}}\right) \Rightarrow CME_i^N = \frac{CME_i}{\sigma_i}.
\end{equation}
where $\sigma_i$ is the $i$-th measurement standard deviation.

$CME$ values are estimated from real and synthetic measurements (SM), after which bad data analysis \cite{bretas2021cyber} is performed. Real measurements are defined as those obtained from sensors. SM are created artificially per the methodology in \cite{synth}. SM are created in low redundancy areas considering measurement $II$ and $n$-tuple of critical measurements, serving to maintain a high global redundancy level $GRL$ (number of measurements divided by the number of state variables). This in turn increases the degrees of freedom such that the $\chi^2$ distribution used for hypothesis testing tends to a normal distribution. SM are calculated based on the previous set of state estimates. The location and type of SM generated are strategically chosen based on the $VI$ of a given bus $i$:

\begin{equation}\label{eq:vi}
VI_i = \sqrt{\frac{1}{L}\sum_{l=1}^{L}\sum_{j=1}^{J}(S_{CME}(l,j)-S_r(l,j)))^2 }
\end{equation}
where $L$ is the set of measurements affiliated with bus $i$, $J$ is the set of all measurements, and $S_{CME}$ and $S_r$ are the sensitivity matrices ($S = 1-K$) of the $CME$ and residual respectively. The $VI$ metric is used to identify buses susceptible to undetectable errors and create SM for those buses.

$\chi^2$ hypothesis testing is used as the GE analytic for bad data detection in the measurement set. The $CME$-based objective function value (\ref{eq:chi-squared}) is compared to a $\chi^2$ threshold, which is based on a chosen probability $p$ (typically $p=0.95$) and the degrees of freedom $d$ of the measurement model:
\begin{equation} \label{eq:chi-squared}
J_{CME}(\hat{\mathbf{x}})=\sum_{i=1}^d \left[\frac{CME_i}{\sigma_i}\right]^2 > \chi^2_{d,p}.
\end{equation}

If the value of $J_{CME}$ is greater than the $\chi^2$ threshold, then a GE is detected. If this occurs, the second SE step is invoked, during which the SE is re-run with meter precision values for each measurement type. Next, following \cite{ParameterProof}, bad data is identified through $CME^N$ analysis. 

It is also important to know which phase 
was affected by the line-to-ground HIF. Thus, a per-phase SE approach is employed \cite{perphase}. In this method, the presented SE is performed, after which the vector of residuals is decomposed into three vectors—one for each phase $A$, $B$, and $C$, i.e., $\Delta \mathbf{z}_A$, $\Delta \mathbf{z}_B$, $\Delta \mathbf{z}_C$. One can then obtain Jacobian matrices for each phase $i$; $\mathbf{H}_i = -\partial\Delta \mathbf{z}_i/\Delta \mathbf{x}_i$. The covariance matrix of the measurements $\mathbf{R}$ is similarly decomposed to obtain per-phase weight matrices $\mathbf{R}^{-1}_i$. The projection matrices for each phase $i$ can then be calculated along with phase analysis quantities $II_i$, $CME_i$, and $CME_i^N$. Accordingly, a respective $J_{CME}(\hat{\mathbf{x}})$ is obtained for each phase, abbreviated as $J_{A}(\hat{\mathbf{x}})$, $J_{B}(\hat{\mathbf{x}})$, and $J_{C}(\hat{\mathbf{x}})$. The $\chi^2$ distribution then uses the number of measurements of each phase $i$ $(mi)$ as degrees of freedom. Thus, the HIF-affected phase $i$ can be identified if $J_{i}(\hat{\mathbf{x}}) > \chi^2_{mi,p}$. 
The SE then communicates with protection systems to locate and isolate the HIF of the affected feeder at relay times of 0.3-0.5 s, however SE could be run locally on relay/recloser phasor measurement units (PMUs), further expediting the process \cite{relay}. 

\subsection{Parameter Error Analysis}\label{sec:paramerror}
To obtain accurate state vector estimates during the SE process, PDS parameters must be pre-established, such as line impedance values. Parameter errors occur when the parameter values used for SE do not match those of the physical system. For this work, the properties of parameter errors are of particular interest. The proposed model, after all, must differentiate between measurement errors and parameter errors caused by the HIF. Measurement errors are identified individually when comparing their $CME^N$ to a threshold $\beta$ (typically equal to three standard deviations of a given measurement, i.e., $\beta$ = 3). If a GE is detected, the $CME^N$s are analyzed. If an isolated measurement is found to have a $CME^N$ above the threshold, it is determined to have a measurement error.

\begin{figure}
\includegraphics[width=7.5072cm,height = 4cm,scale = 1]{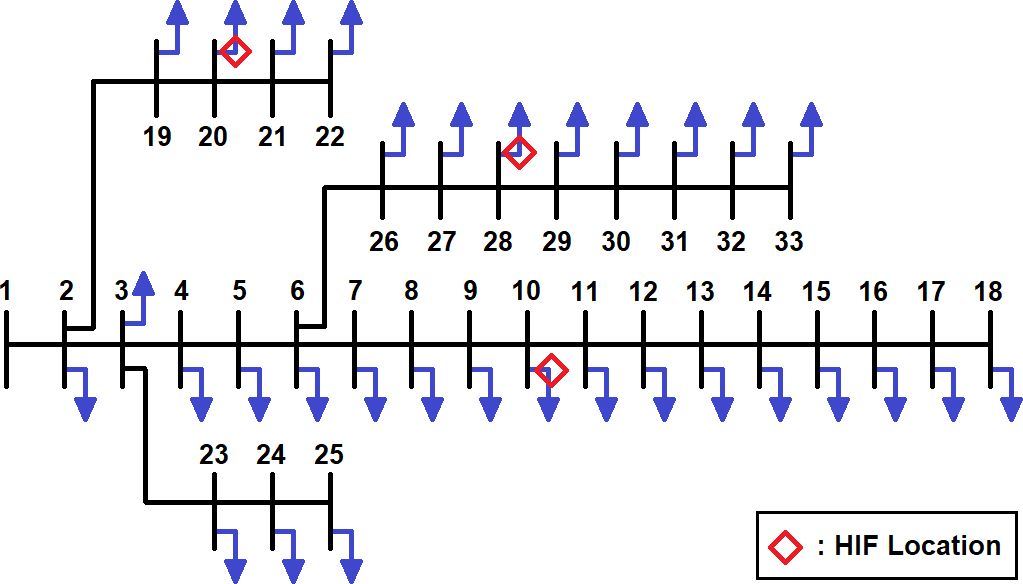}
\centering
\caption{33-Bus Distribution System and HIF Locations}
\label{33bus}
\end{figure}

In contrast to measurement errors, parameter errors spread out to the measurement functions containing the parameter in error, a formal proof of which is presented in \cite{ParameterProof}. Indeed, the measurement model in \eqref{eq:SE} does not consider the possibility of parameter data errors. If one instead considers a correct measurement $z_i$ with a parameter in error $p_i$, one obtains a new measurement model: 
\begin{equation}
    z_i = h(x,p_i) + e_i 
\label{PEModel}\end{equation}

The function model (\ref{PEModel}) can further be developed into a Taylor series form: $z_i = h_{i,0} + \frac{\partial h_i}{\partial p}(x_0,p_0)\Delta p_i$, after which the parameter error $\Delta p_i$ is obtained: $\Delta p_i = \frac{z_i-h_{i,0}}{\mathbf{H}_{p,0}}$, where $\mathbf{H}_{p,0}$ is the Jacobian of parameters. During GE analytics, if it is found that there are measurements $i$-$j$ and $j$-$i$
(real or reactive power flows) and measurements $i$ and $j$ (real or
reactive power injections) with $CME^N$ above the threshold value $\beta$, then the branch $i$-$j$ is suspicious of having a parameter error.

\section{Case Study}\label{casestudy}
\begin{figure}
\includegraphics[width=4.3705cm,height = 3.66445733cm,scale = 1]{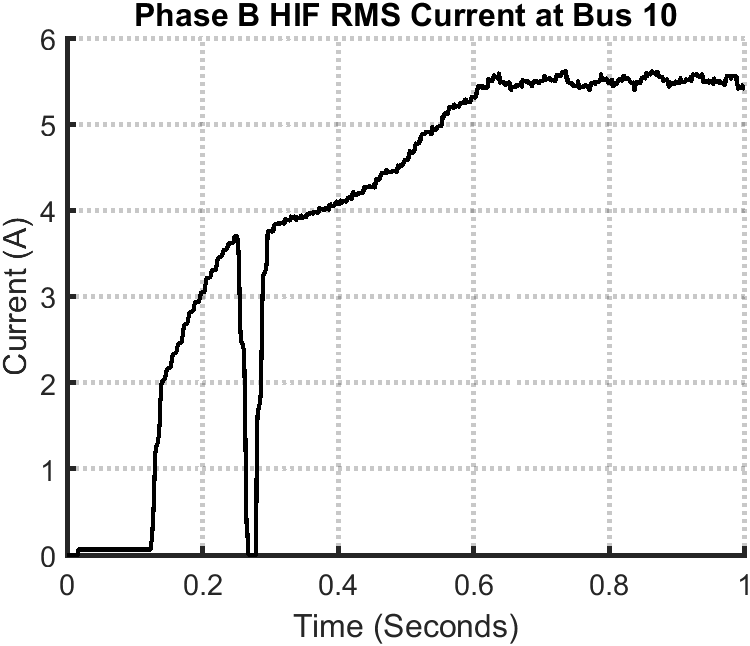}
\includegraphics[width=4.3705cm,height = 3.66445733cm,scale = 1]{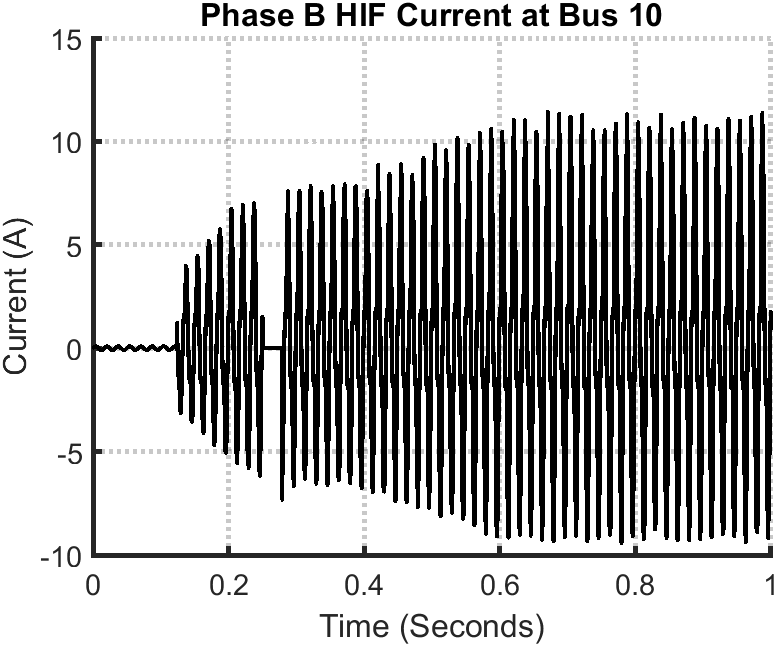}
\includegraphics[width=4.3705cm,height = 3.66445733cm,scale = 1]{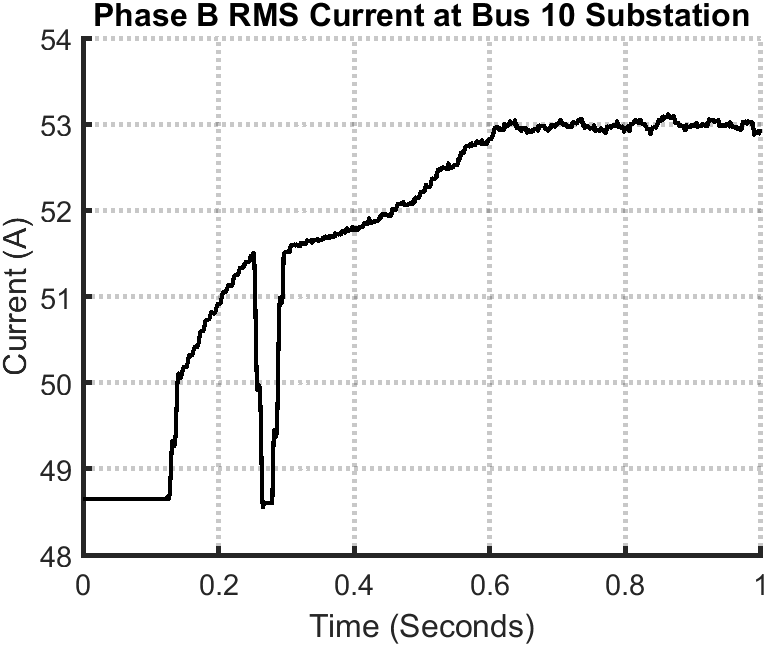}
\centering
\caption{Comparison between HIF and Substation current measurements for a Phase $B$ Fault at the Bus 10 lateral line}
\label{Current Comparisons}
\end{figure}

Validation was performed in Simulink on the 12.66 kV 33-bus distribution system from Baran and Wu \cite{33bus}. Three HIF scenarios are presented. At Bus 20, 10, and 28, phases $A$, $B$, and $C$ respectively are subject to a line-to-ground HIF lasting 1.0 second using the Fig. \ref{HIF Model} model at the lateral line from bus to load. The measurement plan consisted of 195 measurements, leading to $GRL = 3$. A figure of the 33-bus system and proposed HIF locations is included in Fig. \ref{33bus}. To ensure that HIF current is low enough to be categorized as such and that the immediate substation current measurement is low enough so as not to exceed typical overcurrent relay thresholds \cite{overcurrentrelays}, current measurements were recorded, illustrated in Fig. \ref{Current Comparisons} , where substation measurement in this context refers to the recorded current flow from Bus 9 to 10. Tables I, II, and III quantify the low current displacement of the affected phase.

\begin{table}[h]
\caption{Bus 20 Lateral HIF Current Measurements}
\begin{center}
\renewcommand{\arraystretch}{1.5}
\begin{tabular}{ | m{22em} | m{1.2cm}| } 
  \hline
  Phase $A$ I$_{RMS}$ pre-fault & 20.35 A \\ 
  \hline
  Phase $A$ I$_{RMS}$ post-fault when $J_{A}(\hat{\mathbf{x}}) > \chi^2_{195,0.95}$ & 23.73 A \\ 
  \hline
 Phase $A$ I$_{RMS}$ displacement & 3.38 A \\ 
  \hline
  Smallest detectable HIF I$_{RMS}$ & 4.08 A \\ 
  \hline
\end{tabular}
\end{center}
\medskip
\caption{Bus 10 Lateral HIF Current Measurements}
\begin{center}
\renewcommand{\arraystretch}{1.5}
\begin{tabular}{ | m{22em} | m{1.2cm}| } 
  \hline
  Phase $B$ I$_{RMS}$ pre-fault & 48.65 A \\ 
  \hline
  Phase $B$ I$_{RMS}$ post-fault when $J_{B}(\hat{\mathbf{x}}) > \chi^2_{195,0.95}$ & 52.37 A \\ 
  \hline
 Phase $B$ I$_{RMS}$ displacement & 3.72 A \\ 
  \hline
  Smallest detectable HIF I$_{RMS}$ & 4.78 A \\ 
  \hline
\end{tabular}
\end{center}
\medskip
\caption{Bus 28 Lateral HIF Current Measurements}
\begin{center}
\renewcommand{\arraystretch}{1.5}
\begin{tabular}{ | m{22em} | m{1.2cm}| } 
  \hline
  Phase $C$ I$_{RMS}$ pre-fault & 86.07 A \\ 
  \hline
  Phase $C$ I$_{RMS}$ post-fault when $J_{C}(\hat{\mathbf{x}}) > \chi^2_{195,0.95}$ & 89.26 A \\ 
  \hline
 Phase $C$ I$_{RMS}$ displacement & 3.19 A \\ 
  \hline
  Smallest detectable HIF I$_{RMS}$ & 5.92 A \\ 
  \hline
\end{tabular}
\end{center}
\end{table}

Sensor measurements are obtained from the Simulink model, after which the first step of the two-step quasi-static SE is conducted and the first set of SM are generated. Because all measurements are considered as possibly having GEs, each measurement standard deviation is a percentage of the measurement magnitude $(\sigma_i = |z_i|/100)$ \cite{convprop}. Random errors with distribution $X \sim \mathcal{N}(0,1)$ are applied to all measurements.

\begin{figure}
\centering
\includegraphics[width=7.2cm,height = 5.4cm,scale = 1]{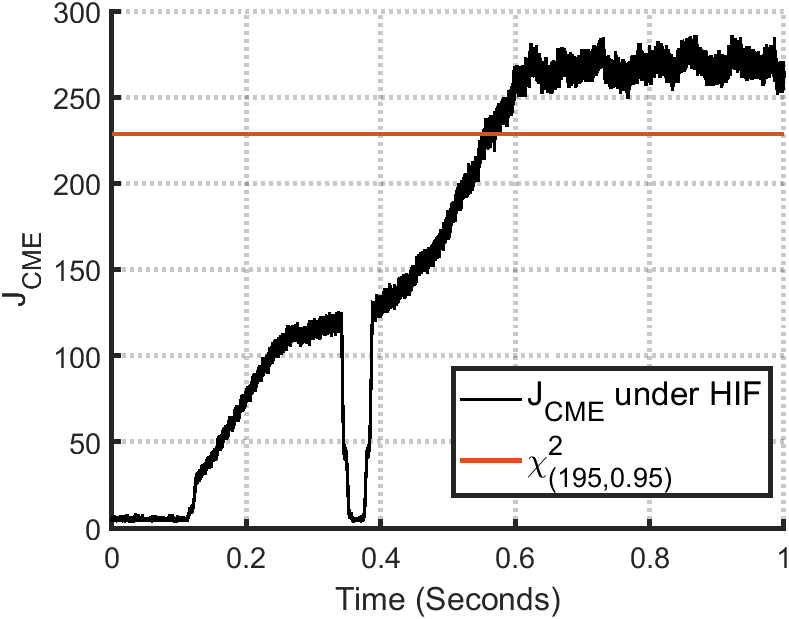}
\caption{$J_{CME}$ evolution for a Phase $C$ lateral HIF at Bus 28}
\label{Jc}
\end{figure}

For a given phase $i$, if $J_{i}(\hat{\mathbf{x}}) > \chi^2_{mi,p}$, a GE is detected. $J_{CME}(\hat{\mathbf{x}})$ evolution for the first SE step during a HIF between Bus 28 and its load is included in Fig. \ref{Jc}; only if $J_{i}(\hat{\mathbf{x}}) > \chi^2_{mi,p}$ does the procedure advance to the second SE step, during which the SE is repeated but with meter precision values for each measurement type. After obtaining a descending order list of $CME^N$, \cite{ParameterProof}, the errors can be classified as either measurement errors or a suspected HIF with parameter error spread. Tables IV, V, and VI catelogue the $CME^N$ distributions, including only those above the detection threshold $\beta = 3$. 

According to the list in Table IV, for example, the phase $A$ injections measurements and associated power flow measurements at Bus 20 presented six $CME^N$ values above the threshold $\beta = 3$, indicative of a HIF at the lateral line from Bus 20 to its load. The highest $CME^N$ corresponds to the real power injection P:20 at the bus of the lateral HIF, followed by the  $i$-$j$ and $j$-$i$ real power flow measurements, P:20-19 and P:19-20, immediately upstream to Bus 20. The next highest error is also associated with Bus 20—the reactive power injection Q:20. Finally, the last set of $CME^N$s above the threshold $\beta = 3$ are the two $i$-$j$ and $j$-$i$ real power flow measurements immediately downstream to Bus 20, P:21-20 and P:20-21. This $CME^N$ distribution is consistent with parameter error spread, as opposed to measurement error \cite{ParameterProof}. Similar measurement function error spread is observed for the Bus 10 and Bus 28 cases (Tables V and VI, respectively). For all three cases, the highest $CME^N$ is the real power injection at the bus of the lateral fault. Monte Carlo simulations were run for the same measurement sets used in Tables IV-VI but with varying sets of noise; the highest $CME^N$ was the respective case's real power injection at the HIF bus location 100\% of the time.

\begin{table}[]
\caption{Bus 20 Lateral HIF $CME^N$ Distribution}
\centering
\renewcommand{\arraystretch}{1.5}
\begin{tabular}{ccc}
\hline
\multicolumn{3}{c}{\begin{tabular}[c]{@{}c@{}}$J_{A}(\hat{\mathbf{x}}) = 250.7226 \ \textgreater \ C = \chi^2_{195,0.95} = 228.5799 \Rightarrow\ $Bad data detected!\end{tabular}} \\ \hline
\multicolumn{3}{c}{$CME^N$ Descending List}                       \\\hline                                                                                      
Measurement                                           & \ \ \ \ \ \ \ \ \ \ \ \ \ \ \ \ \ \ \ \ \ \ \ $II$                                            & \ \ \ \ \ \ \ \ \ \ \ \ \ \ \ \ \ \ \ $CME^N$                                          \\\hline
P:20                                                  & \ \ \ \ \ \ \ \ \ \ \ \ \ \ \ \ \ \ \ \ \ \ \ 1.6388                                          & \ \ \ \ \ \ \ \ \ \ \ \ \ \ \ \ \ 12.5684                                          \\
P:20-19                                               & \ \ \ \ \ \ \ \ \ \ \ \ \ \ \ \ \ \ \ \ \ \ \ 1.4147                                          & \ \ \ \ \ \ \ \ \ \ \ \ \ \ \ \ \ -4.5078                                           \\
P:19-20                                               & \ \ \ \ \ \ \ \ \ \ \ \ \ \ \ \ \ \ \ \ \ \ \ 1.4058                                          & \ \ \ \ \ \ \ \ \ \ \ \ \ \ \ \ \ 4.2587                                          \\
Q:20                                                  & \ \ \ \ \ \ \ \ \ \ \ \ \ \ \ \ \ \ \ \ \ \ \ 1.7335                                          & \ \ \ \ \ \ \ \ \ \ \ \ \ \ \ \ \ -3.9474                                           \\
P:21-20                                               & \ \ \ \ \ \ \ \ \ \ \ \ \ \ \ \ \ \ \ \ \ \ \ 1.1959                                          & \ \ \ \ \ \ \ \ \ \ \ \ \ \ \ \ \ 3.1522                                          \\
P:20-21                                                  & \ \ \ \ \ \ \ \ \ \ \ \ \ \ \ \ \ \ \ \ \ \ \ 1.1954                                          & \ \ \ \ \ \ \ \ \ \ \ \ \ \ \ \ \ -3.0565                                \\\hline         
\end{tabular}
\end{table}

\begin{table}[]
\caption{Bus 10 Lateral HIF $CME^N$ Distribution}
\centering
\renewcommand{\arraystretch}{1.5}
\begin{tabular}{ccc}
\hline
\multicolumn{3}{c}{\begin{tabular}[c]{@{}c@{}}$J_{B}(\hat{\mathbf{x}}) = 241.7247 \ \textgreater \ C = \chi^2_{195,0.95} = 228.5799 \Rightarrow\ $Bad data detected!\end{tabular}} \\ \hline
\multicolumn{3}{c}{$CME^N$ Descending List}                       \\\hline                                                                                      
Measurement                                           & \ \ \ \ \ \ \ \ \ \ \ \ \ \ \ \ \ \ \ \ \ \ \ $II$                                            & \ \ \ \ \ \ \ \ \ \ \ \ \ \ \ \ \ \ \ $CME^N$                                          \\\hline
P:10                                                  & \ \ \ \ \ \ \ \ \ \ \ \ \ \ \ \ \ \ \ \ \ \ \ 1.2862                                          & \ \ \ \ \ \ \ \ \ \ \ \ \ \ \ \ \ 11.7488                                          \\
P:09-10                                               & \ \ \ \ \ \ \ \ \ \ \ \ \ \ \ \ \ \ \ \ \ \ \ 1.5342                                          & \ \ \ \ \ \ \ \ \ \ \ \ \ \ \ \ \ 4.2983                                           \\
P:10-09                                               & \ \ \ \ \ \ \ \ \ \ \ \ \ \ \ \ \ \ \ \ \ \ \ 1.5433                                          & \ \ \ \ \ \ \ \ \ \ \ \ \ \ \ \ \ -3.8723                                          \\
P:11-10                                               & \ \ \ \ \ \ \ \ \ \ \ \ \ \ \ \ \ \ \ \ \ \ \ 1.5546                                          & \ \ \ \ \ \ \ \ \ \ \ \ \ \ \ \ \ 3.3604                                           \\
P:10-11                                               & \ \ \ \ \ \ \ \ \ \ \ \ \ \ \ \ \ \ \ \ \ \ \ 1.5534                                          & \ \ \ \ \ \ \ \ \ \ \ \ \ \ \ \ \ -3.2135                                          \\
P:09                                                  & \ \ \ \ \ \ \ \ \ \ \ \ \ \ \ \ \ \ \ \ \ \ \ 1.0557                                          & \ \ \ \ \ \ \ \ \ \ \ \ \ \ \ \ \ 3.0204                                             \\\hline         
\end{tabular}
\end{table}

\begin{table}[]
\caption{Bus 28 Lateral HIF $CME^N$ Distribution}
\centering
\renewcommand{\arraystretch}{1.5}
\begin{tabular}{ccc}
\hline
\multicolumn{3}{c}{\begin{tabular}[c]{@{}c@{}}$J_{C}(\hat{\mathbf{x}}) = 242.1354$ \ \textgreater \ $C = \chi^2_{195,0.95} = 228.5799 \Rightarrow\ $Bad data detected!\end{tabular}} \\ \hline
\multicolumn{3}{c}{$CME^N$ Descending List}                       \\\hline                                                                                      
Measurement                                           & \ \ \ \ \ \ \ \ \ \ \ \ \ \ \ \ \ \ \ \ \ \ \ $II$                                            & \ \ \ \ \ \ \ \ \ \ \ \ \ \ \ \ \ \ \ $CME^N$                                          \\\hline
P:28                                                  & \ \ \ \ \ \ \ \ \ \ \ \ \ \ \ \ \ \ \ \ \ \ \ 0.9316                                          & \ \ \ \ \ \ \ \ \ \ \ \ \ \ \ \ \ 13.8367                                          \\
P:29-28                                               & \ \ \ \ \ \ \ \ \ \ \ \ \ \ \ \ \ \ \ \ \ \ \ 1.4273                                          & \ \ \ \ \ \ \ \ \ \ \ \ \ \ \ \ \ 6.2406                                           \\
P:27-28                                               & \ \ \ \ \ \ \ \ \ \ \ \ \ \ \ \ \ \ \ \ \ \ \ 1.7423                                          & \ \ \ \ \ \ \ \ \ \ \ \ \ \ \ \ \ 5.7956                                          \\
P:28-29                                               & \ \ \ \ \ \ \ \ \ \ \ \ \ \ \ \ \ \ \ \ \ \ \ 1.4639                                          & \ \ \ \ \ \ \ \ \ \ \ \ \ \ \ \ \ -4.8387                                           \\
P:28-27                                               & \ \ \ \ \ \ \ \ \ \ \ \ \ \ \ \ \ \ \ \ \ \ \ 1.7292                                          & \ \ \ \ \ \ \ \ \ \ \ \ \ \ \ \ \ -4.5188                                          \\
Q:28                                                  & \ \ \ \ \ \ \ \ \ \ \ \ \ \ \ \ \ \ \ \ \ \ \ 0.3208                                          & \ \ \ \ \ \ \ \ \ \ \ \ \ \ \ \ \ -3.7285                                           \\
P:27                                                  & \ \ \ \ \ \ \ \ \ \ \ \ \ \ \ \ \ \ \ \ \ \ \ 0.9033                                          & \ \ \ \ \ \ \ \ \ \ \ \ \ \ \ \ \ 3.7171                                           \\
Q:28                                                  & \ \ \ \ \ \ \ \ \ \ \ \ \ \ \ \ \ \ \ \ \ \ \ 0.4279                                          & \ \ \ \ \ \ \ \ \ \ \ \ \ \ \ \ \ -3.4318                                          \\
P:29                                               & \ \ \ \ \ \ \ \ \ \ \ \ \ \ \ \ \ \ \ \ \ \ \ 2.1124                                          & \ \ \ \ \ \ \ \ \ \ \ \ \ \ \ \ \ 3.1529                                      \\\hline         
\end{tabular}
\end{table}

\section{Summary and Conclusions}\label{sec:conclusion}
This paper presents a parameter error modeling approach for HIF detection under a two-step, per-phase SE framework. Overall, the presented model was successful in detecting low current magnitude HIFs and generating identifiable $CME^N$ error distributions following parameter error methodology. It should be noted that this work does assume synchronized PMU measurements. Sufficient granularity is required to capture the wide-area measurement set needed for the SE process and $\chi^2$ hypothesis test. The first SE step allowed for detection at realistically low HIF currents. The second step revealed HIF parameter error distributions, as opposed to measurement error(s). The error spread of power flow and injection measurements near the HIF, as well as the prominence of errors pertaining to the bus at which the lateral HIF occurred, demonstrate detection and, for future work, location capabilities for HIFs under a parameter error modeling framework.

\bibliographystyle{IEEEtran}
\bibliography{main}

\end{document}